\documentclass[prb,twocolumn,showpacs]{revtex4}
\usepackage[T1]{fontenc}
\usepackage[latin9]{inputenc}
\usepackage{amsmath}
\usepackage{amssymb}
\usepackage{graphicx}
\usepackage{esint}

\makeatother
\begin{document}

\title{False spin zeros in the angular dependence of magnetic quantum
oscillation in quasi-two-dimensional metals}
\author{P.D.~Grigoriev}
\email[Corresponding author; e-mail: ]{grigorev@itp.ac.ru}
\affiliation{L. D. Landau Institute for Theoretical Physics,
142432 Chernogolovka, Russia} \affiliation{National University of
Science and Technology ''MISiS'', Moscow 119049, Russia}
\affiliation{P.N. Lebedev Physical Institute, RAS, 119991, Moscow,
Russia} \affiliation{Institut Laue-Langevin, BP 156, 41 Avenue des
Martyrs, 38042 Grenoble Cedex 9, France}
\author{T.I.~Mogilyuk}
\affiliation{National Research Centre "Kurchatov Institute", Moscow, Russia}

\begin{abstract}
The interplay between angular and quantum magnetoresistance
oscillations in quasi-two-dimensional metals leads to the angular
oscillations of the amplitude of quantum oscillations. This effect
becomes pronounced in high magnetic field, when the simple
factorization of the angular and quantum oscillations is not
valid. The amplitude of quantum magnetoresistance oscillations is
reduced at the Yamaji angles, i.e. at the maxima of the angular
magnetoresistance oscillations. These angular beats of the
amplitude of quantum oscillations resemble and may be confused
with the spin-zero effect, coming from the Zeeman splitting. The
proposed effect of "false spin zeros" becomes stronger in the
presence of incoherent channels of interlayer electron transport
and can be used to separate the different contributions to the
Dingle temperature and to check for violations from the standard
factorization of angular and quantum magnetoresistance
oscillations.
\end{abstract}

\date{\today }

\maketitle

\section{Introduction}

Layered quasi-two-dimensional (Q2D) compounds are of great interest to
modern condensed matter physics and comprise almost all high-temperatures
superconductors, organic metals, intercalated graphites, GaAs layered
heterostructures, rare-earth tellurides and numerous other natural and
artificial layered conductors. The magnetic quantum oscillations (MQO) and
angular dependence of magnetoresistance (MR) are two traditional and common
tools to probe the electronic structure of metals.\cite%
{Shoenberg,Abrik,Ziman} In Q2D metals even the classical MR shows
oscillating behavior as a function of tilt angle $\theta$ of magnetic field
with respect to the normal to conducting layers,\cite{KartsAMRO1988,Yam}
called the angular magnetoresistance oscillations (AMRO). Now, together with
MQO, AMRO are extensively used to study the electronic structure in layered
organic metals (see, e.g., \cite%
{MarkReview2004,Singleton2000Review,KartPeschReview,OMRev,MQORev,Brooks2006,LebedBook}
for reviews), in heterostructures,\cite{Kuraguchi2003} ruthenates,\cite%
{Bergemann2003} tungsten bronze,\cite{AMROBronze} and even in cuprate
high-temperature superconductors.\cite%
{HusseyNature2003,AbdelNature2006,AbdelPRL2007AMRO,McKenzie2007,AMROKartsovnikNd,AMROCuprate2015Analytis}

The Fermi surface in Q2D metals has the shape of a warped cylinder, which
corresponds to the strongly anisotropic electron dispersion
\begin{equation}
\epsilon _{3D}\left( \mathbf{k}\right) \approx \epsilon _{2D}\left(
k_{x},\,k_{y}\right) -2t_{z}\cos (k_{z}d),  \label{ES3D}
\end{equation}%
where $\hbar \left\{ k_{x},\,k_{y},\,k_{z}\right\} $ are the electron
momentum components, $\hbar $ is the Planck's constant, $d$ is the
interlayer distance, and the interlayer transfer integral $t_{z}$ is much
less than the Fermi energy $E_{F}$. In some cases, especially in
low-symmetry crystals, $t_{z}=t_{z}\left( k_{x},\,k_{y}\right) $ depends on
in-plane momentum, which affects AMRO and MQO.\cite%
{Mark92,Singleton2001,Bergemann,GrigAMRO2010} However, to describe most
compounds it is sufficient to take $t_{z}\left( k_{x},\,k_{y}\right) \approx
const$. The geometrical explanation of AMRO\cite{Yam} for the electron
dispersion in Eq. (\ref{ES3D}) is based on the observation that for the
quadratic and isotropic in-plane electron dispersion $\epsilon _{2D}\left(
k_{x},k_{y}\right) =\hbar ^{2}\left( k_{x}^{2}+k_{y}^{2}\right) /2m^{\ast }$
and for $t_{z}\approx const$ the cross-section areas of such
warped-cylindrical Fermi surface in the first order in $t_{z}$ become
independent on $k_{z}$ for some tilt angles $\theta =\theta _{Yam}$ of
magnetic field, now called the Yamaji angles.\cite%
{MarkReview2004,Singleton2000Review,KartPeschReview,OMRev,MQORev,Brooks2006,LebedBook}
The Yamaji angles give the minima of the angular dependence of interlayer
conductivity $\sigma _{zz}\left( \theta \right) $ and correspond to the
zeros of the Bessel function $J_{0}\left( \kappa \right) $, where $\kappa
\equiv k_{F}d\tan \theta $ and $k_{F}$ is the in-plane Fermi momentum. The
direct calculation of interlayer conductivity from the Boltzmann transport
equation in the $\tau $-approximation with the electron dispersion in Eq. (%
\ref{ES3D}) gives\cite{Yagi1990}
\begin{equation}
\frac{\sigma _{zz}\left( \theta \right) }{\sigma _{zz}}=\left[ J_{0}\left(
\kappa \right) \right] ^{2}+2\sum_{\nu =1}^{\infty }\frac{\left[ J_{\nu
}\left( \kappa \right) \right] ^{2}}{1+\left( \nu \omega _{c}\tau \right)
^{2}}\equiv \Phi _{AMRO}\left( \theta \right) ,  \label{sa}
\end{equation}%
where $\tau $ is the electron mean free time, and the cyclotron frequency $%
\omega _{c}$ in Q2D metals depends on the tilt angle $\theta $ of magnetic
field: $\omega _{c}\equiv eB_{z}/m^{\ast }c=\omega _{c0}\cos \theta $, where
$B_{z}$ is the component of magnetic field perpendicular to conducting
layers, $e$ is the electron charge, $m^{\ast }$ is the effective electron
mass and $c$ is the light velocity. In Ref. \cite{Yagi1990} the MQO are
neglected and $\sigma _{zz}\approx \sigma _{zz}^{0}$, where the interlayer
conductivity without magnetic field
\begin{equation}
\sigma _{zz}^{0}=e^{2}\rho _{F}\left\langle v_{z}^{2}\right\rangle \tau
=2e^{2}t_{z}^{2}m^{\ast }\tau d/\pi \hbar ^{4},  \label{s0}
\end{equation}%
$\rho _{F}=m^{\ast }/\pi \hbar ^{2}d$ is the 3D density of states (DoS) at
the Fermi level in the absence of magnetic field per two spin components,
and the mean squared interlayer electron velocity along interlayer direction
is $\left\langle v_{z}^{2}\right\rangle =2t_{z}^{2}d^{2}/\hbar ^{2}$. Eq. (%
\ref{sa}) agrees with the result of Yamaji at $\omega _{c}\tau \rightarrow
\infty $. A microscopic calculation of Q2D AMRO using the Kubo formula and
electron dispersion in Eq. (\ref{ES3D}), neglecting the MQO, also gives Eq. (%
\ref{sa}) when the number of filled Landau levels (LLs) $n_{LL}^{F}\gg 1$.%
\cite{Kur} Assumption $\sigma _{zz}\approx \sigma _{zz}^{0}$ in Eq. (\ref{sa}%
) is valid only in weak magnetic field, such that $\omega _{c}\tau \ll 1$,
so that MQO are negligible and AMRO are also weak. In strong magnetic field,
$\omega _{c}\tau \gtrsim 1$, when both AMRO and MQO are strong, Eq. (\ref{sa}%
) is, generally, incorrect.

The standard theory of MQO and of AMRO considers these two phenomena
independently, i.e. neglecting their interplay, which is valid only in the
limit of weak MQO and AMRO.\cite{Abrik,Ziman} Usually, to analyze the
experimental data in quasi-2D metals in the high-field limit one applies Eq.
(\ref{sa}) with a phenomenological replacement $\sigma _{zz}=\sigma
_{zz}^{MQO}\left( B_{z}\right) $, where $\sigma _{zz}^{MQO}\left(
B_{z}\right) $ depends on magnetic field $\boldsymbol{B}$ only due to the
MQO. Then the angular and field dependence of $\sigma _{zz}\left(
\boldsymbol{B}\right) $ factorize:
\begin{equation}
\sigma _{zz}\left( \boldsymbol{B}\right) =\Phi _{AMRO}\left( \theta \right)
\cdot \sigma _{zz}^{MQO}\left( B\right) ,  \label{FactorS}
\end{equation}%
where $\Phi _{AMRO}\left( \theta \right) $ is given by Eq. (\ref{sa}) and
depends on the field strength $B$ via the product $\omega _{c}\tau $, and $%
\sigma _{zz}=\sigma _{zz}^{MQO}\left( B\right) =\sigma _{zz}^{0}+\tilde{%
\sigma}_{zz}\left( B\right) $ include MQO. The oscillating part $\tilde{%
\sigma}_{zz}$ of conductivity is given by a sum of MQO with all frequencies $%
F^{a}=S_{ext}^{a}\hbar c/2\pi e$, determined by the FS extremal
cross-section areas $S_{ext}^{a}$:\cite{Shoenberg,Abrik,CommentSO,GKM}
\begin{equation}
\frac{\sigma _{zz}}{\sigma _{zz}^{0}}\approx \sum_{a}\frac{g_{0,a}}{g_{tot}}%
\left[ 1+2\sum_{k=1}^{\infty }A_{a}\left( k\right) \cos \left( 2\pi k\frac{%
F^{a}}{B}-\phi _{a}\right) \right] ,  \label{sosc}
\end{equation}%
where the total density of states (DoS) at the Fermi level $%
g_{tot}=\sum_{a}g_{0,a}$ is a sum of the contributions $g_{0,a}$ from all FS
pockets $a$, and the phase shift $\phi _{a}\approx \pi /4$. The MQO
amplitudes $A_{a}\left( k\right) $ depend on the FS geometry, being also
proportional to the product of three damping factors:\cite%
{Shoenberg,Abrik,Ziman} the Dingle factor
\begin{equation}
R_{D}\left( k\right) =\exp \left( \frac{-\pi k}{\omega _{c}\tau }\right) ,
\label{RD}
\end{equation}%
the temperature damping factor
\begin{equation}
R_{T}(k)=\frac{2\pi ^{2}k_{B}Tk/\hbar \omega _{c}}{\sinh \left( 2\pi
^{2}k_{B}Tk/\hbar \omega _{c}\right) },  \label{RT}
\end{equation}%
and the spin factor $R_{S}$, in Q2D metals given by\cite%
{Shoenberg,Abrik,Ziman}
\begin{equation}
R_{S}\left( k\right) =\cos \left( \frac{\pi k\Delta _{Z}}{\hbar \omega _{c}}%
\right) =\cos \left( \frac{\pi gkm^{\ast }}{2m_{e}\cos \theta }\right) ,
\label{Rs}
\end{equation}%
where the Zeeman splitting $\Delta _{Z}=g\hbar eB/2m_{e}c=gB\mu _{B}$ of
electron energy is independent of $\theta $ if the electron $g$-factor $g$
does not depend on $\theta $.\cite{Commentgfactor} In Q2D metals $\hbar
\omega _{c}\propto \cos \theta $, and the spin factor $R_{S}$ results to
strong oscillating angular dependence of the MQO amplitude, given by Eq. (%
\ref{Rs}), which is typical to 2D and Q2D metals and allows
measuring the electron $g$-factor from the so-called \textit{spin
zeros} -- the tilt angles $\theta _{s}$, where the factor in Eq.
(\ref{Rs}) becomes zero.

In Q2D metals with electron dispersion in Eq. (\ref{ES3D}) each FS pocket is
a warped cylinder, giving two FS extremal cross-sections. At $t_{z}\gg \hbar
\omega _{c}$ the difference of these two extremal FS cross-sections areas is
much larger than the LL separation, and one can use the 3D formula in Eq. (%
\ref{sosc}), derived in the lowest order in $\hbar \omega
_{c}/t_{z}$. The simplest (but approximate at $\hbar \omega
_{c}\sim t_{z}$) generalization of Eq. (\ref{sosc}) for $\hbar
\omega _{c}\lesssim t_{z}$, by analogy to the quasi-2D
DoS,\cite{Champel2001} is
\begin{equation}
\frac{\sigma _{zz}}{\sigma _{zz}^{0}}\approx \sum_{a}\frac{g_{0,a}}{g_{tot}}%
\left[ 1+2\sum_{k=1}^{\infty }A_{a}\left( k\right) \cos \left( \frac{2\pi
kF^{a}}{B}\right) \right] ,  \label{soscQ2D}
\end{equation}%
where the amplitudes
\begin{equation}
A_{\alpha }\left( k\right) =\left( -1\right) ^{k}J_{0}\left( \frac{4\pi
kt_{z}}{\hbar \omega _{c}}\right) R_{D}\left( k\right) R_{T}\left( k\right)
R_{S}\left( k\right) ,  \label{Ak}
\end{equation}%
and the summation over $\alpha $ in Eq. (\ref{soscQ2D}) is the summation
over cylindrical FS pockets rather than over FS extremal cross sections $a$
as in Eq. (\ref{sosc}). Note that, contrary to Eq. (\ref{sosc}), in Eq. (\ref%
{soscQ2D}) the phases $\phi _{a}$ are absent; in fact these phases are
contained in the Bessel's functions $J_{0}\left( 4\pi kt_{z}/\hbar \omega
_{c}\right) $ in the amplitudes $A_{\alpha }\left( k\right) $. At $\hbar
\omega _{c}\sim t_{z}$ the higher-order terms in $\hbar \omega _{c}/t_{z}$
become important, and Eqs. (\ref{soscQ2D}) and (\ref{Ak}) modify\cite%
{PhSh,SO,Shub,ChampelMineev} (see, e.g., Eqs. (18)-(21) of Ref. \cite{Shub}
or Eqs. (13)-(15) of Ref. \cite{ChampelMineev}), producing two new physical
effect: the phase shift of beats\cite{PhSh,Shub} and the slow oscillations
of magnetoresistance.\cite{SO,Shub}

When MQO and AMRO are strong, their interplay may become
essential. Then not only Eqs. (\ref{sa}) but also Eq.
(\ref{FactorS}) may be incorrect, i.e. the conductivity is not
simply a product of the AMRO factor in Eq. (\ref{sa}) and the MQO
factor in Eq. (\ref{sosc}) or (\ref{soscQ2D}). Recently, the
influence of strong MQO on the AMRO factor in Eq. (\ref{sa}) was
studied.\cite{TarasPRB2014} It was found that in the high-field
limit $\omega _{c}\gg 1/\tau ,\,t_{z}/\hbar $ the strong MQO
modify AMRO factor in Eq. (\ref{sa}), keeping the AMRO period
almost untouched but changing the AMRO amplitude and its
magnetic-field dependence.\cite{TarasPRB2014} Also the shape of
Landau levels (LL), which is not Lorentzian at $\omega _{c}\gg
1/\tau ,\,t_{z}/\hbar $, is reflected in the AMRO damping. For
example, for Gaussian LL shape the terms with $\nu \neq 0$ are
stronger damped than in Eq. (\ref{sa}) and given by Eq. (33) of
Ref. \cite{TarasPRB2014}, which increases the AMRO amplitude.
Thus, the interplay between MQO and AMRO may be considerable at
$\omega _{c}\tau \gg 1$.

In the present paper we study the influence of AMRO on MQO, especially on
the angular dependence of the amplitude of MQO of magnetoresistance. We show
that this influence is rather strong, and in high magnetic field, at $%
\omega_{c}\gg1/\tau,\,t_{z}/\hbar$, may lead to a new qualitative phenomenon
-- the false spin zeros of MQO of MR.

\section{The model and general formulas}

\subsection{Two-layer model}

To study the influence of AMRO on MQO, we consider strongly anisotropic Q2D
metals in a high magnetic field, when $\omega _{c}\gg 1/\tau ,\,t_{z}/\hbar $
and both AMRO and MQO are strong. In this limit, to calculate the interlayer
conductivity $\sigma _{zz}$ one can apply the two-layer model,\cite%
{MosesMcKenzie1999,WIPRB2011,TarasPRB2014} where $\sigma _{zz}$ is
calculated as a tunnelling conductivity between two adjacent
conducting layers using Kubo formula with electron Green's
function taken inside 2D conducting layer with disorder (see
Appendix A). It was shown that this two-layer model is equivalent
to the 3D models with strongly anisotropic electron dispersion
$\epsilon _{3D}\left( \mathbf{k}\right) $ if $\omega _{c}\gg
1/\tau ,\,t_{z}/\hbar $.\cite{GrigPRB2013} Then
\begin{equation}
\sigma _{zz}\left( T\right) =\frac{1}{2}\sum_{s=\pm 1}\int d\varepsilon %
\left[ -n_{F}^{\prime }(\varepsilon )\right] \sigma _{zz}\left( \varepsilon
+s\Delta _{Z}\right) ,  \label{sT}
\end{equation}%
where $n_{F}^{\prime }(\varepsilon )=-1/\{4T\cosh ^{2}\left[ (\varepsilon
-\mu )/2T\right] \}$ is the derivative of the Fermi distribution function, $%
\mu =E_{F}$ is the chemical potential of electrons, and\cite{TarasPRB2014}
\begin{eqnarray}
\frac{\sigma _{zz}\left( \varepsilon \right) }{\sigma _{zz}^{0}} &=&\frac{%
2\Gamma _{0}\hbar \omega _{c}}{\pi }\sum_{n,p\in Z}Z(n,p)\times  \notag \\
&&\times \text{Im}G(\varepsilon ,\,n)\text{Im}G(\varepsilon ,\,n+p).
\label{I1}
\end{eqnarray}%
Here the function $Z$ comes from the overlap of electron wave functions on
adjacent layers, producing AMRO, and is given by Eq. (12) of Ref. \cite%
{TarasPRB2014}, which coincides with the square of Eq. (9) in Ref. \cite%
{Raikh}. The interlayer conductivity in the absence of magnetic field is
given by $\sigma _{zz}^{0}=2e^{2}\tau _{0}m^{\ast }t_{z}^{2}d/\pi \hbar ^{4}$%
, and $\Gamma _{0}=\hbar /2\tau _{0}$.
Eq. (\ref{I1}) is valid for arbitrary electron Green's functions
\begin{equation}
G(\varepsilon ,\,n)=\frac{1}{\varepsilon -\hbar \omega _{c}(n+1/2)-\Sigma
(\varepsilon )},  \label{Grn}
\end{equation}%
which contain the self-energy part $\Sigma (\varepsilon )$ determined by
disorder. Below we neglect the electron-electron (e-e) interaction, which
can be used only when many LL are filled, $n_{LL}^{F}=\left\lceil \mu /\hbar
\omega _{c}\right\rceil \gg 1$, and the e-e interaction is effectively
screened.\cite{KukushkinUFN1988,Aleiner1995} In this limit $n_{LL}^{F}\gg 1$
the function $Z(n,p)$ simlifies\cite{TarasPRB2014,Kur}
\begin{equation}
Z(n,p)\approx Z(n_{LL}^{F},p)\approx J_{p}^{2}\left( k_{F}d\tan \theta
\right) \equiv J_{p}^{2}\left( \kappa \right) .  \label{Z2}
\end{equation}

With notations $\Gamma (\varepsilon )=\left\vert \text{{Im}}\Sigma
\left( \varepsilon \right) \right\vert =-${Im}$\Sigma ^{R}\left(
\varepsilon \right) $ and $\varepsilon ^{\ast }\equiv \varepsilon
-$Re$\Sigma (\varepsilon )$, the imaginary part of the electron
Green's function is
\begin{equation}
\text{Im}G(\varepsilon ,\,n)=-\Gamma (\varepsilon )/\left[
(\varepsilon ^{\ast }-\hbar \omega
_{c}(n+1/2))^{2}+\Gamma^{2}(\varepsilon )\right] ,  \label{Lor}
\end{equation}%
Using also the notations $\varepsilon ^{\ast }\equiv \varepsilon
-$Re$\Sigma (\varepsilon )$, $\gamma _{0}=2\pi \Gamma _{0}/\hbar
\omega _{c}$, $\gamma \equiv 2\pi \left\vert \text{Im}\Sigma
(\varepsilon )\right\vert /\hbar
\omega _{c}$, and $\alpha \equiv 2\pi \varepsilon ^{\ast }/\hbar \omega _{c}$%
, from Eqs. (\ref{I1})-(\ref{Lor}) one obtains
\begin{equation}
\frac{\sigma _{zz}(\varepsilon )}{\sigma _{zz}^{0}}=\frac{\Gamma _{0}}{%
\Gamma (\varepsilon )}\sum_{p=-\infty }^{\infty }S_{p}\left[
J_{p}\left( \kappa \right) \right] ^{2},  \label{s1L1}
\end{equation}%
where
\begin{eqnarray}
S_{0} &\equiv &\sum_{n\in Z}\frac{(2/\pi )\hbar \omega _{c}\Gamma ^{3}%
}{\left[ \left( \varepsilon ^{\ast }-\hbar \omega _{c}\left( n+\frac{1}{2}%
\right) \right) ^{2}+\Gamma ^{2}\right] ^{2}}=  \label{S0} \\
&=&\frac{\sinh (\gamma )}{\cos (\alpha )+\cosh (\gamma )}-\gamma \frac{%
1+\cos (\alpha )\cosh (\gamma )}{(\cos (\alpha )+\cosh (\gamma ))^{2}}
\notag
\end{eqnarray}%
in agreement with Eq. (23) of Ref. \cite{ChampelMineev}, and for $p\neq 0$
\begin{gather}
S_{p}\equiv \sum_{n\in Z}\frac{\left( 2/\pi \right) \hbar \omega
_{c} \Gamma ^{3}}{\left[ \left( \varepsilon ^{\ast }-\hbar \omega
_{c}\left(
n+1/2\right) \right) ^{2}+\Gamma^{2}\right] }\times  \notag \\
\times \frac{1}{\left[ \left( \varepsilon ^{\ast }-\hbar \omega _{c}\left(
n+p+1/2\right) \right) ^{2}+\Gamma^{2}\right] }=  \notag \\
=\frac{\sinh (\gamma )}{\left( \cos (\alpha )+\cosh (\gamma )\right) \left[
1+(p\pi /\gamma )^{2}\right] }.  \label{Sp}
\end{gather}%
Equations (\ref{s1L1})-(\ref{Sp}) give both AMRO and MQO for arbitrary
(unknown yet) electron self-energy $\Sigma \left( \varepsilon \right) $.

\begin{figure}[htb]
\centering\includegraphics[width=1\linewidth]{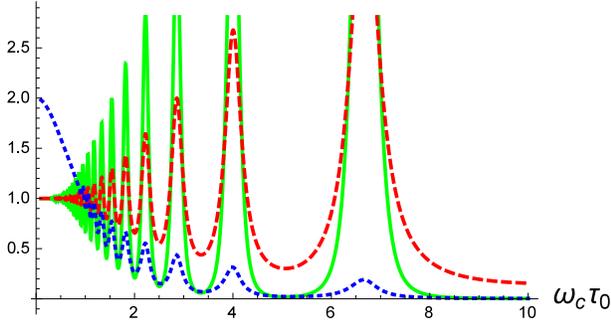}
\caption{(Color on-line) The comparison of the functions $S_{0}\left(
B_{z}\right) $ (solid green line) and $S_{p}\left( B_{z}\right) $ at $p=0$
(dashed red line), given by Eqs. (\protect\ref{S0}) and (\protect\ref{Sp})
at $\left\vert \text{{Im}}\Sigma \left( \protect\varepsilon \right)
\right\vert =\Gamma _{0}=const$ and $\protect\alpha =20\protect\gamma $.
Both functions show oscillations around the same background, but the
amplitude of the oscillations of $S_{0}\left( B_{z}\right) $ are much
stronger at $\protect\gamma \ll 1$. The dotted blue line gives $2S_{p}\left(
p=1\right) $ for comparison. }
\label{FigS}
\end{figure}

At $\gamma \gg 1$ (weak field limit) the second term is Eq. (\ref{S0}) is
exponentially small, so that $S_{0}$ in Eq. (\ref{S0}) is the same as $S_{p}$
in Eq. (\ref{Sp}) at $p=0$. Hence, as expected, at $\gamma \gg 1$ we confirm
Eqs. (\ref{sa}) and (\ref{FactorS}). However, at $\gamma \ll 1$ (high-field
limit) the second term in Eq. (\ref{S0}) is important, and the function $%
S_{0}$ in Eq. (\ref{S0}) becomes completely different from $S_{p}(p=0)$ in
Eq. (\ref{Sp}). This means that at $\gamma \ll 1$ Eqs. (\ref{sa}) and (\ref%
{FactorS}) are not valid for any self-energy part $\Sigma \left(
\varepsilon \right) $. The difference between the functions $S_{0}$ and $%
S_{p}(p=0)$, leadinig to the violation of Eqs. (\ref{sa}) and (\ref{FactorS}%
), is illustrated in Fig. \ref{FigS} at $\left\vert \text{{Im}}\Sigma \left(
\varepsilon \right) \right\vert =\Gamma _{0}=const$ and is clearly seen
already from the expansions of $S_{0}$ and of $S_{p}(p=0)$ at $\gamma
\rightarrow 0$:
\begin{gather}
S_{0}\left( \gamma \rightarrow 0\right) =\frac{2-\cos (\alpha )}{[1+\cos
(\alpha )]^{2}}\frac{\gamma ^{3}}{3}+O[\gamma ]^{5},  \label{Sexp} \\
S_{p}\left\vert _{\substack{ \gamma \rightarrow 0,  \\ p=0}}\right. =\frac{%
\gamma }{1+\cos (\alpha )}-\frac{2-\cos (\alpha )}{[1+\cos (\alpha )]^{2}}%
\frac{\gamma ^{3}}{6}+O[\gamma ]^{5}.  \notag
\end{gather}%
For example, in the minima of MQO of conductivity, i.e. at $\alpha =0$, Eq. (%
\ref{Sexp}) shows that the function $S_{0}$ is much smaller than $S_{p}(p=0)$
at $\gamma \rightarrow 0$:%
\begin{equation}
S_{0}\left\vert _{\substack{ \gamma \rightarrow 0,  \\ \alpha =0}}\right.
\approx \frac{\gamma ^{3}}{12}\ll S_{p}\left\vert _{\substack{ \gamma
\rightarrow 0,  \\ p=0  \\ \alpha =0}}\right. \approx \frac{\gamma }{2}.
\label{Sexp1}
\end{equation}%
These expansions (\ref{Sexp}) and (\ref{Sexp1}) are valid at any $\alpha $
except the proximity of the point $\alpha =\pi $, where the denominator in (%
\ref{Sexp}) vanishes. At $\alpha =\pi $ and $\gamma \rightarrow 0$ the
expansion of Eqs. (\ref{S0}) and (\ref{Sp}) gives%
\begin{equation}
S_{0}\left\vert _{\substack{ \gamma \rightarrow 0,  \\ \alpha =\pi }}\right.
\approx \frac{4}{\gamma },~~S_{p}\left\vert _{\substack{ \gamma \rightarrow
0,  \\ p=0  \\ \alpha =\pi }}\right. \approx \frac{2}{\gamma },
\end{equation}%
i.e. in the maxima of MQO at $\gamma \rightarrow 0$ the function $S_{0}$ is
only twice larger than $S_{p}(p=0)$.

As was shown in Refs. \cite{Shub,ChampelMineev}, the function $S_{0}$
differs from $S_{p}(p=0)$ because of the extra term $G_{R}^{2}$ in the Kubo
formula for conductivity. This term contributes only a second-order poles in
the integrand over $\varepsilon $, which does not affect the result at zero
magnetic field but contributes the term $\sim $ $\gamma $ to the amplitudes
of MQO harmonics of conductivity.\cite{Shub,ChampelMineev}

\subsection{Electron Green's function and self energy}

The expression (\ref{Grn}) for the electron Green's function contains the
self-energy $\Sigma \left( \varepsilon \right) $, which at low temperature
mainly comes from the scattering on impurity potential
\begin{equation}
V_{i}\left( \mathbf{r}\right) =\sum_{j}U\delta ^{3}\left( \mathbf{r}-\mathbf{%
r}_{j}\right) .  \label{Vi}
\end{equation}%
The impurities are assumed to be short-range (point-like) and randomly
distributed with volume concentration $n_{i}$. The scattering by this
impurity potential is spin-independent. In the non-crossing (self-consistent
single-site) approximation the electron self-energy satisfies the following
equation:\cite{Ando}
\begin{equation}
\Sigma (\varepsilon )=\frac{n_{i}U}{1-UG\left( \varepsilon \right) },
\label{SER}
\end{equation}%
where the averaged Green's function in the coinciding points\cite%
{ChampelMineev,Grigbro}
\begin{gather}
G\left( \varepsilon \right) =\sum_{n,k_{y},k_{z}}G\left( \varepsilon
,n\right) =\frac{g_{LL}}{d}\sum_{n=0}^{+\infty }G\left( \varepsilon
,n\right)   \label{Gr0} \\
\approx \frac{g_{LL}}{d}\sum_{n=-\infty }^{+\infty }\frac{1}{\varepsilon
-\hbar \omega _{c}(n+1/2)-\Sigma (\varepsilon )}  \label{Gr1} \\
=-\frac{\pi g_{LL}}{\hbar \omega _{c}d}\tan \left[ \pi \frac{\varepsilon
-\Sigma (\varepsilon )}{\hbar \omega _{c}}\right] .  \label{Gr}
\end{gather}%
The summation over $k_{y}$ in Eq. (\ref{Gr0}) gives the LL degeneracy $%
g_{LL}=eB_{z}/2\pi \hbar c$, and the summation over $k_{z}$ gives
$1/d$. Strictly speaking, in Eqs. (\ref{Gr0}) the summation over
$n$ must be cut at $n_{\max }\sim W/\hbar \omega _{c}$, where
$W\sim \mu $ is the band width, as the expression logarithmically
diverges. Similarly, in Eq. (\ref{Gr1}) we extended the summation
over $n$ from $-\infty $, because the neglected difference
$\sum_{n=-\infty }^{0}G\left( \varepsilon ,n\right) \approx \ln
\left( W/\mu \right) /\hbar \omega _{c}=const$ does not affect
observable quantities. In the self-consistent Born approximation
(SCBA), used below, the neglected difference is equivalent to the
constant shift of chemical potential.

It is convenient to use the normalized electron Green's function
\begin{equation}
g\left( \varepsilon \right) \equiv G\left( \varepsilon \right) \hbar \omega
_{c}d/\pi g_{LL}.  \label{gnorm}
\end{equation}%
To obtain the monotonic growth of longitudinal interlayer magnetoresistance%
\cite{WIPRB2011,GrigPRB2013,WIJETP2011} and other qualitative
physical effects\cite{TarasPRB2014}, the SCBA is sufficient, which
instead of Eq. (\ref{SER}) gives
\begin{equation}
\Sigma (\varepsilon )-n_{i}U=n_{i}U^{2}G\left( \varepsilon \right) =\Gamma
_{0}g\left( \varepsilon \right) .  \label{SigSCBA}
\end{equation}%
Here we used that the zero-field level broadening is $\Gamma _{0}=\pi
n_{i}U^{2}\nu _{3D}=\pi n_{i}U^{2}g_{LL}/(d\hbar \omega _{c})=\hbar /2\tau
_{0}$. Below we also neglect the constant energy shift $n_{i}U$ in Eq. (\ref%
{SigSCBA}), which does not affect physical quantities as conductivity.

Eqs. (\ref{Gr})-(\ref{SigSCBA}) give the equations on the Green's function $%
g\equiv g\left( \varepsilon \right) $:
\begin{equation}
\text{Im}g=\frac{\sinh \left( \gamma _{0}\text{Im}g\right) }{\cosh \left(
\gamma _{0}\text{Im}g\right) +\cos \left( \alpha \right) }=\frac{\gamma }{%
\gamma _{0}},  \label{Img}
\end{equation}%
\begin{equation}
\text{Re}g=\frac{-\sin \left( \alpha \right) }{\cosh \left( \gamma _{0}\text{%
Im}g\right) +\cos \left( \alpha \right) },  \label{Reg}
\end{equation}%
or the equations for the electron self-energy $\Sigma ^{R}(\varepsilon )$:
\begin{equation}
\frac{\gamma }{\gamma _{0}}=\frac{\sinh \left( \gamma \right) }{\cosh \left(
\gamma \right) +\cos \left( \alpha \right) },  \label{gamma}
\end{equation}%
\begin{equation}
\delta \equiv \alpha -\frac{2\pi \varepsilon }{\hbar \omega _{c}}=\frac{%
\gamma _{0}\sin \left( \alpha \right) }{\cosh \left( \gamma \right) +\cos
\left( \alpha \right) }.  \label{alpha}
\end{equation}%
Here we have used the notations introduced after Eq. (\ref{Lor}). The
solution of Eq. (\ref{gamma}) gives Im$\Sigma (\alpha )$, while Eq. (\ref%
{alpha}) allows to find $\alpha \left( \varepsilon \right) $ and Re$\Sigma
(\varepsilon )$. The system of Eqs. (\ref{gamma}) and (\ref{alpha}) differs
from Eq. (30) of Ref. \cite{ChampelMineev} even in the absence of electron
reservoir (at $R=0$), because in Eq. (30) of Ref. \cite{ChampelMineev} the
oscillating real part of the electron self energy is neglected, which leads
to a different dependence of $\sigma _{zz}(B_{z})$.\cite{GrigPRB2013}

Eq. (\ref{gamma}) allows to find the value $\gamma _{0c}$, when the LLs
become isolated in SCBA, i.e. when the DoS and Im$\Sigma ^{R}(\varepsilon )$
between LLs become zero. In the middle between two adjacent LLs $\cos \left(
\alpha \right) =1$, and equation (\ref{gamma}) for $\gamma _{min}$ at
conductivity minima becomes:
\begin{equation}
\frac{\gamma _{min}}{\gamma _{0}}=\frac{\sinh \left( \gamma _{min}\right) }{%
\cosh \left( \gamma _{min}\right) +1}=\tanh \left( \gamma _{min}/2\right) .
\label{EqGm}
\end{equation}%
This equation always has a trivial solution $\gamma =0$. However, at $\gamma
_{0}>\gamma _{0c}=2$, corresponding to $\pi \Gamma _{0}>\hbar \omega _{c}$,
Eq. (\ref{EqGm}) also has a non-zero solution. This nonzero solution means a
finite DoS at energy between LLs, i.e. at $\pi \Gamma _{0}<\hbar \omega _{c}$
in SCBA\ the LLs become isolated, which affects physical observables, e.g.
leads to a monotonic growth of $\sigma _{zz}(B_{z})$.\cite{Grigbro} \newline

In middle of LL $\cos \left( \alpha \right) =-1$, and equation (\ref{gamma})
for $\gamma _{max}$ at conductivity maxima becomes:
\begin{equation}
\frac{\gamma _{max}}{\gamma _{0}}=\frac{\sinh \left( \gamma _{max}\right) }{%
\cosh \left( \gamma _{max}\right) -1}=\coth \left( \gamma _{max}/2\right) .
\label{EqGmax}
\end{equation}%
This equation always has a non-zero solution.

Any additional Fermi-surface parts, which are not responsible for
the given MQO, create an extra density of states (DoS) at the
Fermi level. This additional DoS does not oscillate with the same
frequency and acts as an electron
reservoir,\cite{ChampelMineev,GrigChemPotOsc,ChampelChemPotOsc}
smearing the MQO. This additional DoS does not oscillate at all if
it comes from open Fermi-surface parts. In this case Eq.
(\ref{gamma}) modifies to
\begin{equation}
\frac{\gamma }{\gamma _{0}}=\left( \frac{\sinh \left( \gamma \right) }{\cosh
\left( \gamma \right) +\cos \left( \alpha \right) }+R\right) /\left(
1+R\right) ,  \label{gammaR}
\end{equation}%
similar to Eq. (24) of Ref. \cite{ChampelMineev}, where $R$ is the
ratio of the reservoir DoS to the average DoS on the Fermi-surface
pocket responsible for MQO.

\section{Interplay between angular and quantum magnetic oscillations}

The influence of MQO on AMRO was already studied
recently.\cite{TarasPRB2014} In this section we analyze the
influence of AMRO on MQO of interlayer conductivity using the
formulas in the previous section II. As it was shown in Sec. IIA,
the violation of Eqs. (\ref{sa}) and (\ref{FactorS}) and new
interesting effects appear only in the high-field limit $\gamma
\ll 1$ and only because of the difference between the functions
$S_{0}$ and $S_{p}\left( p=0\right) $ given by Eqs. (\ref{S0}) and
(\ref{Sp}).\ In this section we consider two limiting cases: (i)
the limit of large electron reservoir, when $\gamma \approx
const$, and (ii) the limit of zero electron reservoir, when there
are no other Fermi-surface pockets except the one responsible for
MQO.

\begin{figure}[htb]
\centering\includegraphics[width=1\linewidth]{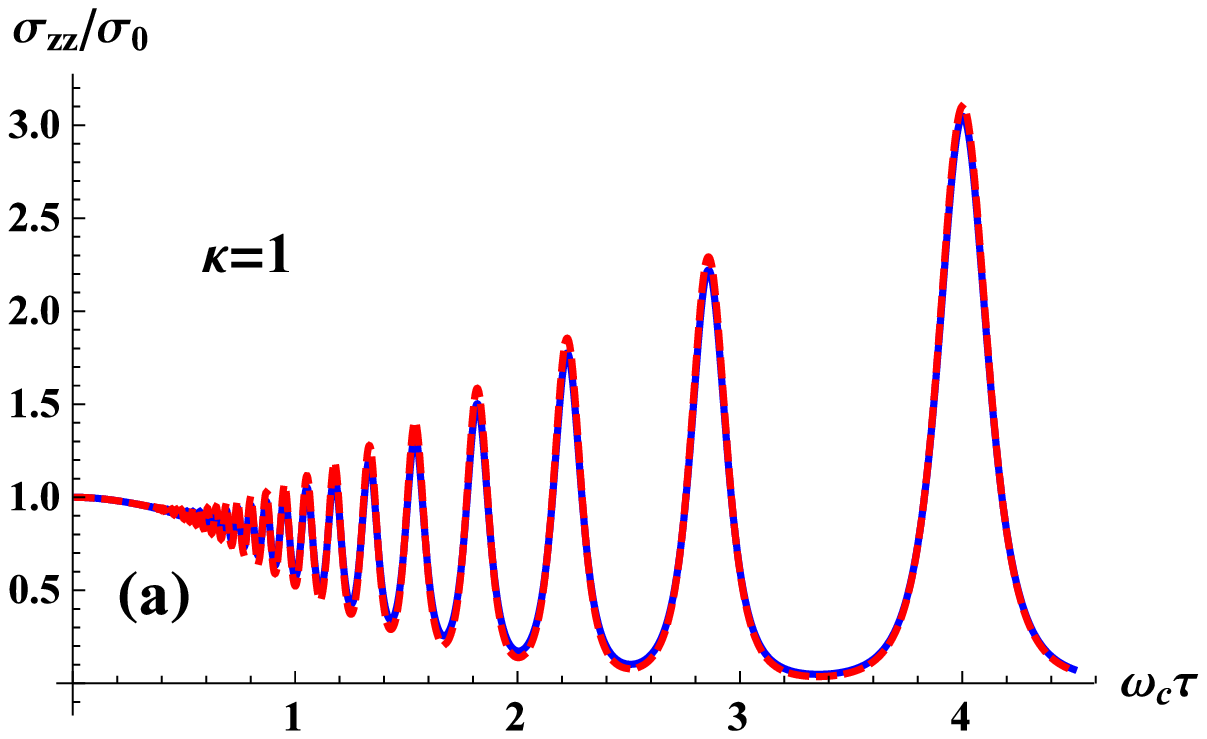}\newline
\bigskip \includegraphics[width=1\linewidth]{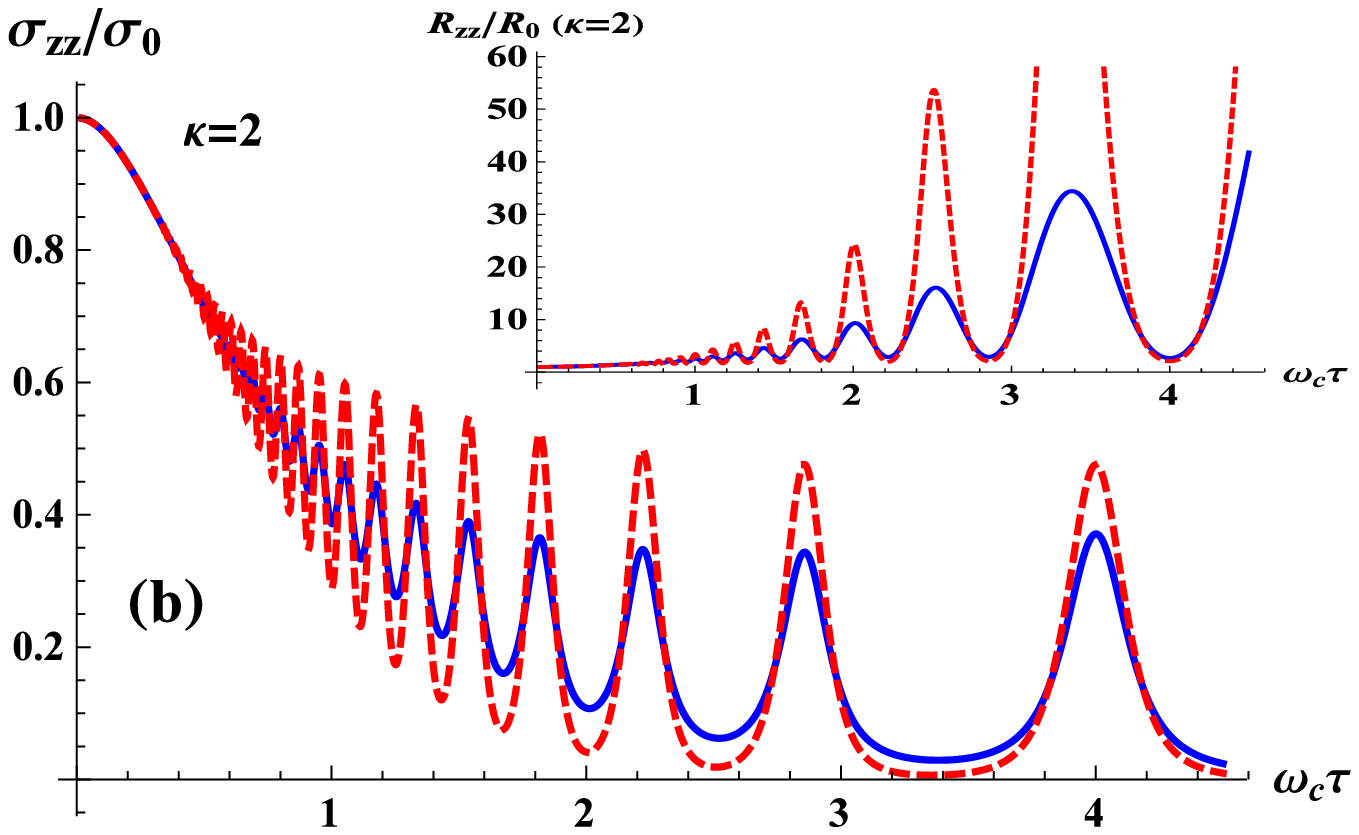}\newline
\bigskip \includegraphics[width=1\linewidth]{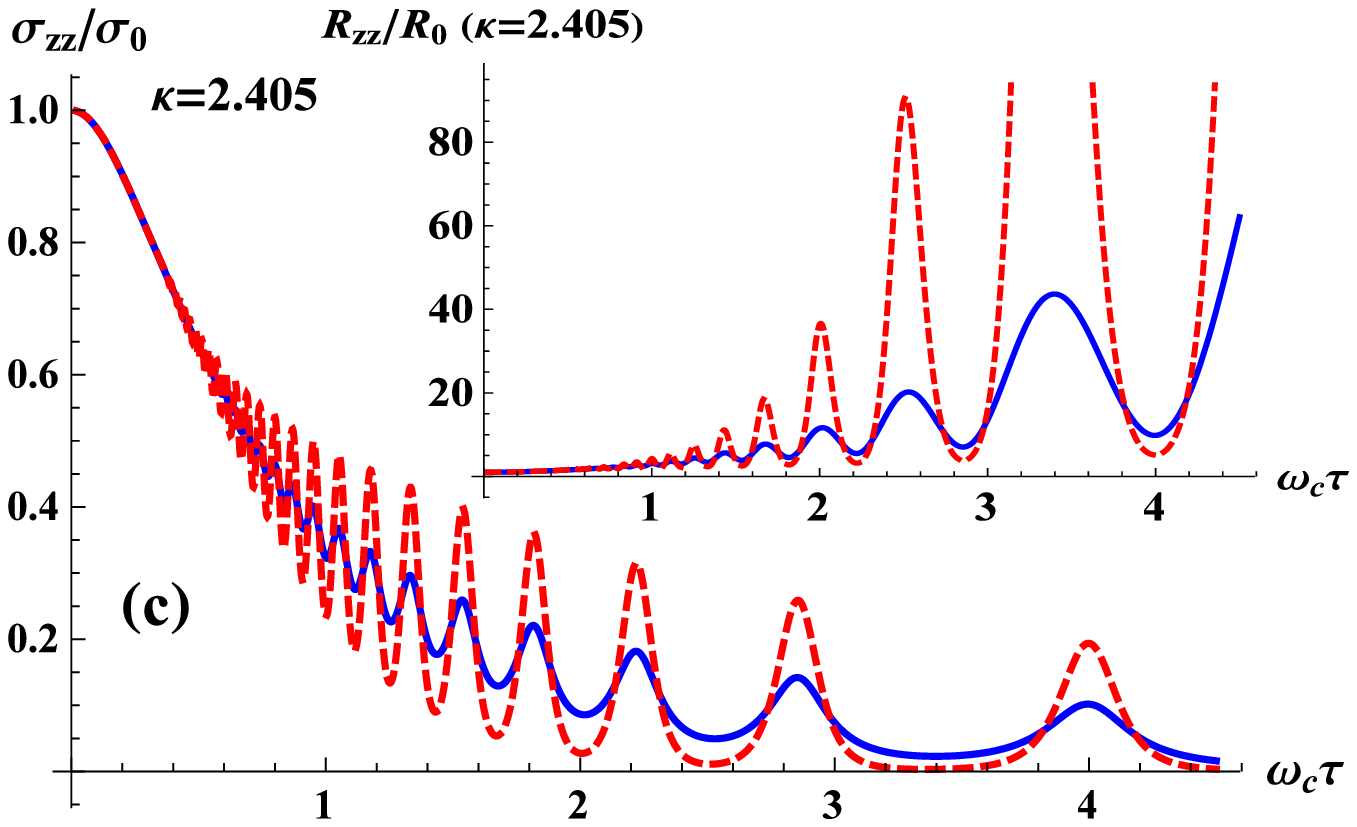}
\caption{(Color on-line) The comparison of the new formula for conductivity $%
\protect\sigma _{zz}^{new}(B_{z})$, given by Eqs. (\protect\ref{s1L1})-(%
\protect\ref{Sp}) and shown by solid blue line, with $\protect\sigma %
_{zz}^{old}(B_{z})$, given by Eqs. (\protect\ref{FactorS}),(\protect\ref{sa}%
) with $\protect\sigma _{zz}^{MQO}\left( B\right) =\protect\sigma %
_{zz}^{0}S_{0}\left( B_{z}\right) $ and shown by dashed red line. For this
comparison we take three different values of $\protect\kappa \equiv
k_{F}d\tan \protect\theta $: $\protect\kappa _{1}=1$, $\protect\kappa _{2}=2$%
, and $\protect\kappa _{3}=2.405$. The latter corresponds to the
first Yamaji angle. The inserts show resistivity
$R_{zz}(B_{z})\approx 1/\protect\sigma _{zz}(B_{z})$.}
\label{FigSig1}
\end{figure}

\subsection{Limit of large electron reservoir and $\left\vert \text{{Im}}%
\Sigma \left( \protect\varepsilon \right) \right\vert \approx const$}

The violation of Eqs. (\ref{sa}) and (\ref{FactorS}) should be strongest in
the minima and maxima of conductivity MQO, where the the functions $S_{0}$
and $S_{p}\left( p=0\right) $ are most different (see Fig. \ref{FigS}).
Additionally, the violation of Eqs. (\ref{sa}) and (\ref{FactorS}) is
expected to be most evident near the Yamaji angles, where the term with $p=0$
in Eq. (\ref{s1L1}) is reduced as compared to the terms with $p\neq 0$.

To check how strong are these deviations from Eqs. (\ref{sa}) and (\ref%
{FactorS}) at $\left\vert \text{{Im}}\Sigma \left( \varepsilon \right)
\right\vert \approx const$, in Fig. \ref{FigSig1} we compare $\sigma
_{zz}^{new}(\varepsilon )$ calculated using Eqs. (\ref{s1L1})-(\ref{Sp}) and
$\sigma _{zz}^{old}(\varepsilon )$ calculated using Eqs. (\ref{FactorS}),(%
\ref{sa}) and Eq. (23) of Ref. \cite{ChampelMineev}, i.e. $\sigma
_{zz}^{MQO}\left( B\right) =\sigma _{zz}^{0}S_{0}$. From this comparison one
sees that indeed the notable violation from Eqs. (\ref{sa}) and (\ref%
{FactorS}) appears at $\left\vert \text{{Im}}\Sigma \left(
\varepsilon \right) \right\vert \approx const$ only near the
Yamaji angles. These deviations do not change the frequency or the
phase of MQO, but considerably reduce their amplitude. This
decrease of MQO amplitude near the Yamaji angles as compared to
the prediction of Eqs. (\ref{sa}) and (\ref{FactorS}) is even more
clear on the magnetoresistance $R_{zz}\approx 1/\sigma _{zz}$,
shown in the inserts to Fig. \ref{FigSig1}. Our result that in the
Yamaji angles the MQO amplitude decreases contradicts the general
opinion that the magnetoresistance oscillations should be stronger
in the Yamaji angles because the system becomes effectively
two-dimensional. Fig. \ref{FigSig1} also illustrate a strong
influence of AMRO on the amplitude of MQO.

\begin{figure}[tbph]
\centering\includegraphics[width=1\linewidth]{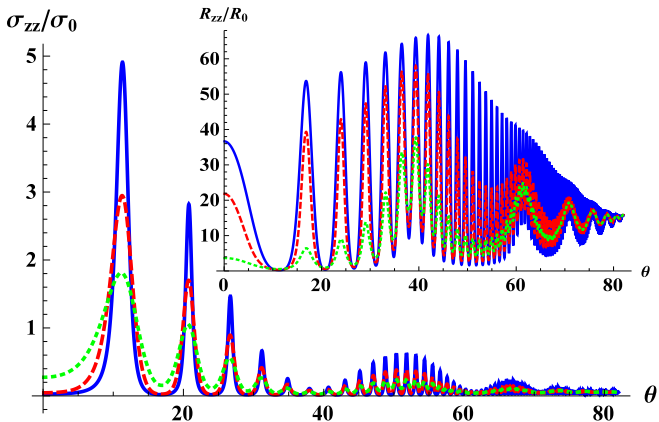}
\caption{(Color on-line) The angular dependence of conductivity (main
figure) and magnetoresistance (insert figure) at $k_Fd=3$, $\protect\omega %
_{c0}\protect\tau _{0}=5$ and at three temperatures: $T=0.1\Gamma_0$ (blue
solid line), $T=0.5\Gamma_0$ (red dashed line), and $T=\Gamma_0$ (green
dotted line). The minima of MQO amplitude, arising from the influence of
AMRO on MQO, may be erroneously treated as spin-zeros.}
\label{FigTAng}
\end{figure}

The angular dependence of conductivity and of magnetoresistance as
a functions of the tilt angle $\theta $ for a constant magnetic
field strength $B_{0}$, calculated using Eqs.
(\ref{sT}),(\ref{s1L1})-(\ref{Sp}), are plotted in Fig.
\ref{FigTAng} at two temperatures: $T=0.1\Gamma_0$ (blue solid
line) and at $T=0.4\Gamma_0$ (red dashed line). $E_F=201\Gamma_0$
and $\omega _{c0}\tau _{0}=5$ at $\theta =0$. The fast quantum
oscillations come from the angular dependence of normal-to-layer
component $B_{z}=B_{0}\cos \theta $ of magnetic field, which
enters the MQO. According to the above analytical estimates, the
amplitude of MQO considerably decreases near the Yamaji angles,
which in Fig. \ref{FigTAng} is seen as the angular oscillations of
the amplitude of MQO. In the analysis of experimental data on
magnetoresistance such beats of the MQO amplitude may be
mistakenly interpreted as spin zeros. We suggest the name
\textit{false spin zeros} for this phenomenon of the angular beats
of MQO amplitude due to the interplay between AMRO and MQO in
quasi-2D metals. Increasing of temperature damps the MQO, but
these \textquotedblright false spin zeros\textquotedblright\ are
still visible.

\begin{figure}[tbph]
\includegraphics[width=1\linewidth]{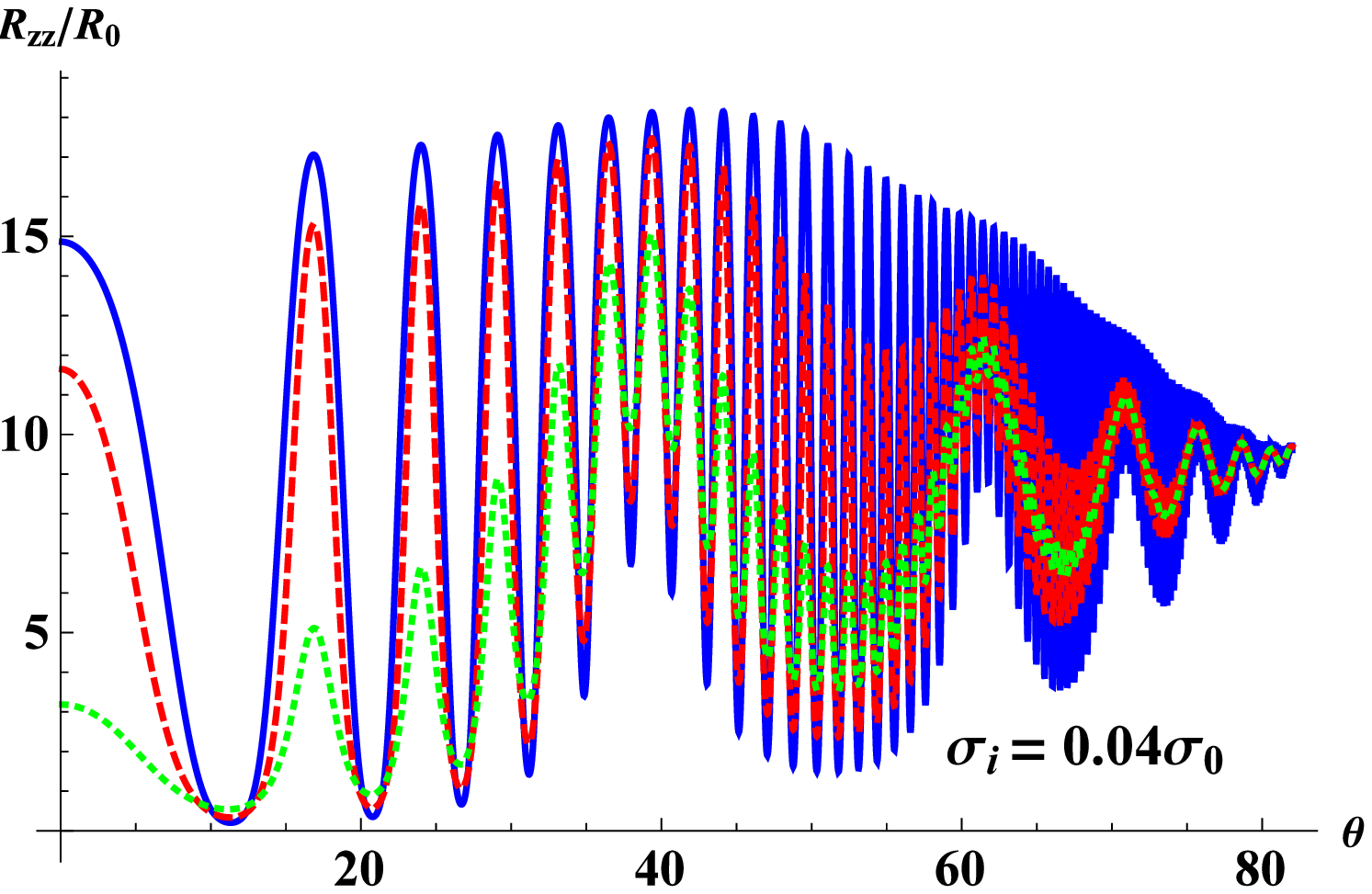} \includegraphics[width=1%
\linewidth]{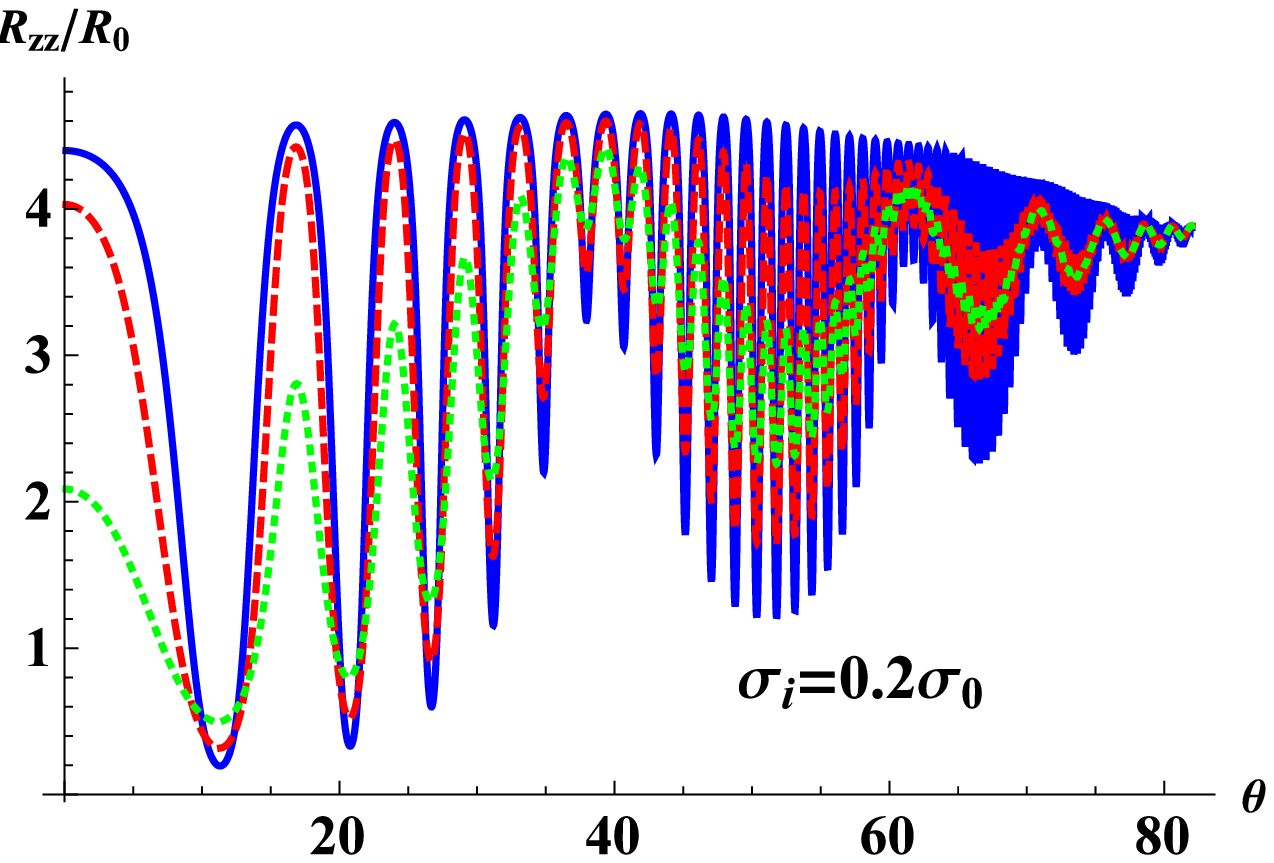} \caption{(Color on-line) The angular dependence
of magnetoresistance in the
presence of incoherent channel of interlayer conductivity at $\protect\sigma%
^{i}_{zz}=0.04\protect\sigma_{0}$ (Fig. a) and $\protect\sigma^{i}_{zz}=0.2%
\protect\sigma_{0}$ (Fig. b) for three temperatures $T=0.1\Gamma_0$ (blue
solid line), $T=0.5\Gamma_0$ (red dashed line), and $T=\Gamma_0$ (green
dotted line). }
\label{FigTAngInc}
\end{figure}

The false spin zeros become even more pronounced if one takes into account
the incoherent channels of interlayer conductivity, which come from crystal
imperfections, from resonance impurities between the conducting layers,\cite%
{Abrikosov1999,Maslov,Incoh2009} or from polaron tunnelling\cite%
{Lundin2003,Ho}. The incoherent channels produce additional term $%
\sigma^{i}_{zz}$ for the interlayer conductivity. This term has neither
angular nor quantum oscillations and shifts conductivity in Fig. \ref%
{FigTAng} upward by a constant. The total conductivity is a sum of the
coherent and incoherent conductivity channels: $\sigma^{tot}_{zz}=\sigma
_{zz}^{coh}+\sigma^{i}_{zz}$. Usually, in clean metals the ratio $%
\sigma^{i}_{zz}/\sigma_0\ll 1$. In Fig. \ref{FigTAngInc} we plot the angular
dependence of interlayer magnetoresistance $R_{zz}=1/\sigma^{tot}_{zz}$ for
two different values of this ratio: $\sigma^{i}_{zz}/\sigma_0=0.04$ (Fig. %
\ref{FigTAngInc}a) and $\sigma^{i}_{zz}/\sigma_0=0.2$ (Fig. \ref{FigTAngInc}%
b). The magnetic field strength in Fig. \ref{FigTAngInc} corresponds to $%
\omega _{c0}\tau _{0}=5$ at $\theta =0$, and $k_Fd=3$. The false
spin zeros, seen as the angular beats of MQO amplitude, in Fig.
\ref{FigTAngInc} are clearer than in Fig. \ref{FigTAng}.

The long-range disorder, which have the length scale greater than
the magnetic length, affects the MQO amplitude differently from
the short-range disorder.\cite{Raikh} The macroscopic sample
inhomogeneities locally shift the Fermi level and damp the MQO
similar to the temperature smearing of the Fermi level. However,
this type of disorder keeps the AMRO amplitude unchanged, similar
to the amplitude of the so-called slow oscillations of
magnetoresistance.\cite{SO,Shub} Using these slow oscillations in
organic metals it was shown\cite{SO} that the contribution
$T_{D}^{inh}$ of such sample inhomogeneities to the total
Dingle temperature $T_{D}$ of MQO exceeds more than four times the contribution $%
T_{D}^{*}$ to the Dingle temperature from the short-range
disorder. This information is helpful to understand the nature of
disorder in various compounds. The observation of slow
oscillations requires that the Landau-level separation
$\hbar\omega_c$ is less than the interlayer transfer
integral $t_z$ but exceeds the Dingle temperature, i.e. $t_z>\hbar\omega_c>%
\Gamma_0$. In very anisotropic compounds, where $t_z\lesssim
\Gamma_0$, this condition cannot be satisfied, and the slow
oscillations are very difficult to observe. However, just in this
limiting case the comparison of the amplitudes of AMRO and MQO
allows to determine the contribution $T_{D}^{inh}$ of these sample
inhomogeneities from experimental data on magnetoresistance,
because the amplitude of AMRO is not affected by $T_{D}^{inh}$
contrary to the amplitude of MQO. The observation of false spin
zeros in the amplitude of MQO and their temperature evolution
increases the accuracy of such extraction of various contributions
to the Dingle temperature from experimental data.

\begin{figure}[tbph]
\centering\includegraphics[width=\linewidth]{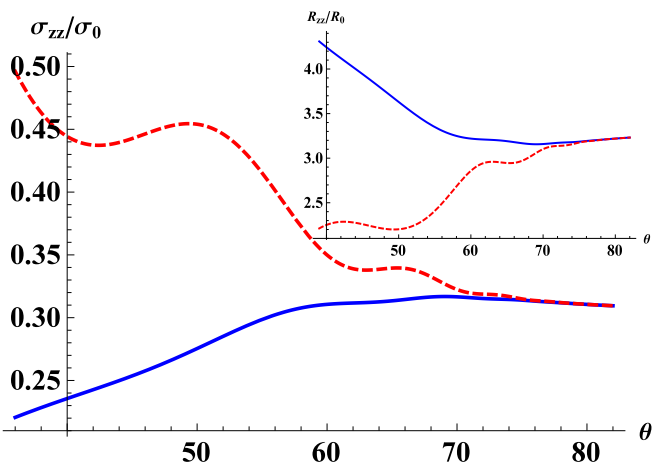}
\caption{(Color on-line) The maxima (solid blue line) and minima
(dashed red line) of quantum oscillations of interlayer
conductivity (main panel) and of magnetoresistance (insert figure)
calculated in SCBA in the absence of electron reservoir as
function of the tilt angle $\theta $ of magnetic field at
$k_{F}d=3$ and $\protect\gamma _{0}=2$. These plots give the
angular dependence of the envelope of MQO.} \label{FigEnv}
\end{figure}

\subsection{Conductivity in the absence of electron reservoir}

\begin{figure}[tbph]
\includegraphics[width=0.5\textwidth]{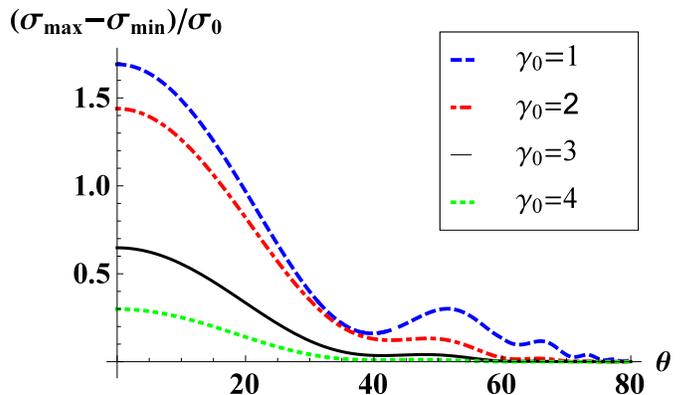}
\caption{(Color on-line) Angular dependence of relative amplitude $\left( \protect\sigma %
_{zz}^{\max }-\protect\sigma _{zz}^{\min }\right) /\protect\sigma
_{0}$ of
the oscillations of interlayer conductivity at $k_{F}d=3$ for various $%
\protect\gamma _{0}$ calculated in SCBA in the absence of electron
reservoir.} \label{FigSCBA1}
\end{figure}

\begin{figure}[tbph]
\includegraphics[width=0.5\textwidth]{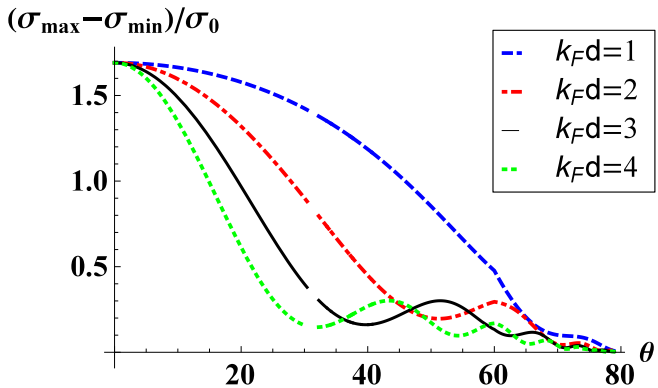}
\caption{(Color on-line) Angular dependence of relative amplitude $\left( \protect\sigma %
_{zz}^{\max }-\protect\sigma _{zz}^{\min }\right) /\protect\sigma %
_{0}$ of oscillations of interlayer conductivity for various $%
k_{F}d $ at $\protect\gamma _{0}=3$ calculated in SCBA.}
\label{FigSCBA2}
\end{figure}

In the absence of electron reservoir the electron Green's function
and self-energy are given by Eqs. (\ref{Img})-(\ref{alpha}). In
this limit, to calculate conductivity one needs to solve the
self-consistency equation (\ref{gamma}) for the self energy, which
can be done only numerically. In the minima and maxima of MQO of
conductivity Eq. (\ref{gamma}) simplifies to Eqs. (\ref{EqGm}) and
(\ref{EqGmax}) correspondingly, which are convenient to calculate
the envelope of MQO, shown in Fig. \ref{FigEnv} for $k_{F}d=3$ and
$\gamma _{0}=\pi /\omega _{c}\tau _{0}=2$. In Fig. \ref{FigSCBA1}
we plot the normalized amplitude of MQO of conductivity $\left(
\sigma _{zz}^{\max }-\sigma _{zz}^{\min }\right) /\sigma _{0}$ for
$k_{F}d=3$ and for various values of $\gamma _{0}$. In Fig.
\ref{FigSCBA2} we plot the normalized amplitude of MQO of
conductivity $\left( \sigma _{zz}^{\max }-\sigma _{zz}^{\min
}\right) /\sigma _{0}$ for $\gamma _{0}=3$ and for various values
of $k_{F}d$. These plots show that the false spin zeros are more
pronounced at larger $k_{F}d$, when AMRO are faster, and are
easier observed at $\gamma _{0}<4$. Note, that in both limiting
cases, i.e. at large and at zero electron reservoir, the proposed
"false spin zeros" only decrease the amplitude of MQO but do not
produce the phase inversion of MQO. Thus, contrary to the true
spin zeros, given by the factor $R_{s}$ in Eq. (\ref{Rs}), which
changes the sign and thus leads to the phase shift of MQO by $\pi
$, the false spin zeros are not strong enough to make such
inversion of MQO. This difference can be used to distinguish
between the true and the false spin zeros on experimental data.



\section{Conclusions}

In this paper we analyze the influence of the angular oscillations
of magnetoresistance (AMRO) in quasi-2D metals on its quantum
oscillations. We show that the previous assumption of
factorization of these two types of oscillations, given by Eq.
\ref{FactorS} and usually applied to analyze experimental data,
violates in high magnetic field when $\omega _{c}\tau \gtrsim 1$.
The strongest violation of Eq. \ref{FactorS} is near the Yamaji
angles (AMRO maxima), where the amplitude of MQO is strongly
reduced. This interplay of AMRO and MQO at $\omega _{c}\tau
\gtrsim 1$ leads to the new qualitative effect -- the oscillations
(beats) of the amplitude of MQO as function of the tilt angle
$\theta $ of magnetic field. These angular minima of MQO
amplitude, originating from AMRO and called "false spin zeros",
may be erroneously treated as true spin zeros and lead to the
incorrect determination of the electron g-factor from MQO. The
proposed false spin zeros do not produce the phase inversion of
MQO and thus can be distinguished from the true spin zeros. The
false spin zeros are more pronounced at larger values of $\omega
_{c}\tau $ (see Fig. \ref{FigSCBA1}) and at larger values of
$k_Fd$, when AMRO have larger frequency (see Fig. \ref{FigSCBA2}).
The incoherent channels of interlayer conductivity also make the
proposed effect of "false spin zeros" stronger, which is seen from
the comparison shown in Figs. \ref{FigTAng} and \ref{FigTAngInc}.
The false spin zeros may help to determine the contribution of
such incoherent channel to the total interlayer conductivity from
experimental data. The comparison of the amplitude of angular and
quantum oscillations may help to determine the nature of disorder
which contributes to the Dingle temperature.

\acknowledgments{The authors are grateful to T.~Ziman for useful
discussions. PG thanks RSCF \#16-42-01100 and TM thanks RFBR
\#16-02-0052 for financial support.}

\appendix

\section{The two-layer model}

In this appendix we remind the formulation and basic formulas of the
two-layer model for interlayer conductivity, developed in Refs. \cite%
{TarasPRB2014,MosesMcKenzie1999,WIPRB2011}. The one-electron Hamiltonian in
layered metals with small interlayer coupling consists of the 3 terms%
\begin{equation}
\hat{H}=\hat{H}_{0}+\hat{H}_{t}+\hat{H}_{I}.  \label{H}
\end{equation}%
The first term $\hat{H}_{0}$ is the 2D free electron Hamiltonian summed over
all layers $j$ and all quantum numbers $\left\{ m\right\} $ of electrons in
magnetic field on a 2D conducting layer:%
\begin{equation*}
\hat{H}_{0}=\sum_{m,j}\epsilon _{2D}\left( m\right) c_{m,j}^{+}c_{m,j},
\end{equation*}%
where $\epsilon _{2D}\left( m\right) =\epsilon _{n}=\hbar \omega _{c}\left(
n+1/2\right) $\ is the corresponding free electron dispersion, and $%
c_{m}^{+}(c_{m})$ are the electron creation (annihilation) operators in the
state $\left\{ m\right\} $. The second term in Eq. (\ref{H}) gives the
coherent electron tunnelling between two adjacent layers:
\begin{equation}
\hat{H}_{t}=2t_{z}\sum_{j}\int d^{2}{\boldsymbol{r}}[\Psi _{j}^{\dagger }({%
\boldsymbol{r}})\Psi _{j-1}({\boldsymbol{r}})+\Psi _{j-1}^{\dagger }({%
\boldsymbol{r}})\Psi _{j}({\boldsymbol{r}})],  \label{Ht}
\end{equation}%
where $\Psi _{j}(x,y)$ and $\Psi _{j}^{\dagger }(x,y)$\ are the creation
(annihilation) operators of an electron on the layer $j$ at the point $(x,y)$%
. This interlayer tunnelling Hamiltonian is called "coherent" because it
conserves the in-plane electron momentum during the interlayer tunnelling.
The last term $\hat{H}_{I}$ is the impurity potential $V_{i}\left( \mathbf{r}%
\right) $, e.g. given by Eq. (\ref{Vi}).

In the strongly anisotropic almost 2D limit, $t_{z}\ll \Gamma ,\hbar \omega
_{c}$, the interlayer hopping $t_{z}$ can be considered as a perturbation
for the periodic stack of uncoupled 2D metallic layers. The interlayer
conductivity $\sigma _{zz}$, associated with the Hamiltonian (\ref{Ht}), can
be calculated using the Kubo formula as a tunnelling conductivity between
two adjacent conducting layers $j$ and $j+1$: \cite%
{MosesMcKenzie1999,WIPRB2011}%
\begin{eqnarray}
\sigma _{zz} &=&\frac{4e^{2}t_{z}^{2}d}{\pi \hbar }\int d\varepsilon \left[
-n_{F}^{\prime }(\varepsilon )\right] \int d^{2}{\boldsymbol{r}}\left\langle
ImG({\boldsymbol{r}},j,\varepsilon ) \right\rangle  \notag \\
&&\times \left\langle ImG(-{\boldsymbol{r}},j+1,\varepsilon )\right\rangle ,
\label{KuboA}
\end{eqnarray}%
where the electron Green's function $G({\boldsymbol{r}},j,\varepsilon )$ on
the metallic layer $j$ includes the scattering by impurities. The angular
brackets in Eq. (\ref{KuboA}) mean averaging over impurity configurations.
Assuming the impurity distributions on adjacent layer are uncorrelated, the
impurity averaging for each Green's function in Eq. (\ref{KuboA}) is
performed independently.\cite{CommentAv} Then the averaged Green's function
depends only on the difference ${\boldsymbol{r}}$ of the two coordinates: $%
\left\langle G({\boldsymbol{r}}_{1},{\boldsymbol{r}}_{2},j,\varepsilon
)\right\rangle =\left\langle G({\boldsymbol{r}},j,\varepsilon )\right\rangle
$, where ${\boldsymbol{r}}={\boldsymbol{r}}_{2}-{\boldsymbol{r}}_{1}$.

The AMRO of interlayer conductivity in Eq. (\ref{KuboA}) appear because in a
magnetic field $B=\left( B_{x},0,B_{z}\right) =\left( B\sin \theta ,0,B\cos
\theta \right) $, tilted by the angle $\theta $ to the normal to conducting
layers, the Green's functions on two adjacent layers acquire the phase shift
(see Eq. (49) of Ref. \cite{MosesMcKenzie1999}):
\begin{equation}
G_{R}({\boldsymbol{r}},j+1,\varepsilon )=G_{R}({\boldsymbol{r}}%
,j,\varepsilon )\exp \left\{ ie\Lambda \left( {\boldsymbol{r}}\right) /\hbar
\right\} ,  \label{GL}
\end{equation}%
where
\begin{equation}
\Lambda \left( {\boldsymbol{r}}\right) =-yB_{x}d=-yBd\sin \theta .
\label{LambdaPhSh}
\end{equation}%
In the so-called \textquotedblright non-crossing\textquotedblright\
approximation, where the electron self-energy contains only diagrams without
intersections of impurity lines, the averaged Green's function on each layer
factorizes to (see Appendix in Ref.~\cite{WIPRB2011} for the proof)
\begin{equation}
G({\boldsymbol{r}}_{1},{\boldsymbol{r}}_{2},\varepsilon )=\sum_{n,k_{y}}\Psi
_{n,k_{y}}^{0\ast }(\boldsymbol{r}_{2})\Psi _{n,k_{y}}^{0}(\boldsymbol{r}%
_{1})G\left( \varepsilon ,n\right) .  \label{Gg}
\end{equation}%
Then the integration over ${\boldsymbol{r}}$ in Eq. (\ref{KuboA}) can be
performed analytically and gives\cite{TarasPRB2014} Eqs. (\ref{I1}) and (\ref%
{Z2}).


\begin{thebibliography}{99}
\bibitem{CommentSO} In Q2D metals with dispersion in Eq. (\ref{ES3D}) and $
t_{z}\sim\hbar\omega_{c}$ the interlayer conductivity $\sigma_{zz}$ in
addition to MQO has the so-called slow oscillations\cite%
{SO,Shub,GKM,SinchenkoSO} with frequency
$F_{slow}=4t_{z}m^{\ast}c/e\hbar$ determined by  interlayer
hopping integral $t_{z}$ rather than FS cross-section area.
\bibitem{Commentgfactor} The electron $g$-factor in metals may differ from
the free-electron $g$-factor $g\approx 2$ because of electron interaction,
and in non-magnetic compounds $g$-factor is almost independent of $\theta $.
\bibitem{CommentAv} The separate averaging of the spectral functions in Eq.
( \ref{KuboA}) also assumes the neglection of vertex corrections\cite{Mahan}
in the Kubo formula. The vertex corrections to the interlayer conductivity
in Eq. (\ref{KuboA}) are negligibly small because they contain the product
of the electron wave functions on adjacent layers, which is small by the
factor $\sim t_{z}/E_{F}\ll 1$. In normal 3D metals the vertex corrections
on point-like impurities also vanish; they lead to the replacement of the
mean scattering time $\tau $ by the transport mean scattering time,\cite%
{Mahan} and these two times coincide for scattering on point-like
impurities.
\end{thebibliography}
\end{document}